\documentclass[aps,pre,twocolumn,showpacs,superscriptaddress,amsmath,amssymb]{revtex4}

\usepackage{graphicx}
\usepackage{dcolumn}
\usepackage{bm}
\usepackage{color}
\usepackage{multirow}
\usepackage{textcomp}
\usepackage{umoline}

\begin{document}

\title{Phase-shift inversion in oscillator systems with periodically switching couplings}

\author{Sang Hoon Lee}
\email[Corresponding author: ]{sanghoon.lee@physics.umu.se}
\affiliation{IceLab, Department of Physics,
Ume{\aa} University, 901 87 Ume{\aa}, Sweden}

\author{Sungmin Lee}
\affiliation{IceLab, Department of Physics,
Ume{\aa} University, 901 87 Ume{\aa}, Sweden}
\affiliation{Department of Biotechnology,
Norwegian University of Science and Technolgy,
N-7491 Trondheim, Norway}

\author{Seung-Woo Son}
\affiliation{Department of Applied Physics, Hanyang University, Ansan 426--791, Korea}

\author{Petter Holme}
\affiliation{IceLab, Department of Physics,
Ume{\aa} University, 901 87 Ume{\aa}, Sweden}
\affiliation{Department of Energy Science, Sungkyunkwan University, Suwon 440--746, Korea}

\date{\today}

\begin{abstract}
A system's response to external periodic changes can provide crucial information
about its dynamical properties. We investigate the synchronization transition, an archetypical
example of a dynamic phase transition, in the framework of such a temporal response.
The Kuramoto model under periodically switching interactions has the same type of phase
transition as the original mean-field model. Furthermore, we see that the signature
of the synchronization transition appears in the relative delay of the order parameter
with respect to the phase of oscillating interactions as well. Specifically, the phase
shift becomes significantly larger as the system gets closer to the phase transition so that
the order parameter at the minimum interaction density can even be larger than
that at the maximum interaction density, counterintuitively.
We argue that this phase-shift inversion
is caused by the diverging relaxation time, in a similar way
to the resonance near the critical point in the kinetic Ising model. Our result, based on
exhaustive simulations on globally coupled systems as well as scale-free networks,
shows that an oscillator system's phase transition can be manifested in the temporal response
to the topological dynamics of the underlying connection structure.
\end{abstract}



\maketitle

\section{introduction}
A collection of interacting coupled oscillators is one of the most intensively studied systems showing a dynamic phase
transition called a synchronization transition~\cite{KuramotoBook,PikovskyBook,Acebron2005}.
Such systems have been used as models to describe various phenomena such as the pacemakers of the heart,
the collection of amoeba, large-scale ecosystems~\cite{WinfreeBook,KJLee1996,Ivanov1996,Blasius1999,Kori2004,Radicchi2006},
and fiber-optic networks of optoelectronic oscillators~\cite{Ravoori2011}.
As the coupling strength increases, such systems undergo a phase transition from
the desynchronized (disordered) state to the synchronized (ordered) state
characterized by critical phenomena.
The most well-known theoretical formalism is the Kuramoto
model~\cite{KuramotoBook} where the oscillators interact with each other by
a sinusoidal coupling strength with respect to the phase differences of the two
oscillators.

Aside from the steady-state behavior after the initial transient behavior,
the dynamical aspect of the oscillators has also been studied recently~\cite{Strogatz1991,SWSon2008}.
Especially, the effects of temporally varying interaction structures~\cite{Holme2011} themselves
are worthwhile to investigate since we can systematically analyze the response of systems
to such structural changes of interactions~\cite{Belykh2004,Stilwell2006}. For instance,
the mobile oscillator is considered in Refs.~\cite{Frasca2008,Peruani2010,Fujiwara2011}
as an example of temporally switching interactions. In this Brief Report, we take
the Kuramoto model with periodically switching interactions and analyze the
temporal response in the collective phase of oscillators to such periodicity in
interactions.
Such periodically switching interactions may
have implications to biological or social communications among compartments
under natural circadian rhythms, for instance~\cite{CircadianRhythm}.

In terms of the temporally averaged order parameter and its amplitude of oscillation,
we verify that the model undergoes the same synchronization transition as the
mean-field (MF) Kuramoto model~\cite{Acebron2005} with the finite-size scaling (FSS) analysis.
Furthermore, with numerical simulations and analogy to the kinetic Ising model~\cite{BJKim2001},
we show that the signature of the synchronization transition is revealed in the relative
phase shift of the order parameter with respect to the density of interactions, as a resonance-like phase-shift
inversion. The result is also consistent with the diverging relaxation time
at the critical coupling strength~\cite{Strogatz1991,SWSon2008}.
Therefore, we claim that the system's temporal response to the temporally changing interactions can
be used as an indicator of a phase transition.

\section{model}
\label{model}
We consider the Kuramoto-type oscillator dynamics~\cite{Acebron2005} composed of
$N$ oscillators as
\begin{equation}
\frac{d\phi_i}{dt} = \omega_i + \frac{2K}{N} \sum_{j=1}^{N} a_{ij} T_{ij}(t) \sin(\phi_i - \phi_j) ,
\label{Kuramoto_equation}
\end{equation}
where $\{ \phi_i \}$ is the set of oscillators' phases, $\{ \omega_i \}$ is the set of
natural frequencies of oscillators given by the Gaussian distribution with average $0$ and unit
variance, and the symmetric adjacency matrix $a_{ij} = 1$ if $i$ and $j$ are connected and $0$ otherwise.
The interaction $a_{ij} T_{ij}(t)$ mediated by the edge between the oscillators $i$ and $j$
is subject to the periodic
function as
$T_{ij}(t) = H\left[ \sin(\Omega t + \theta_{ij} ) \right]$,
where $H$ is the Heaviside step function, $\Omega$ is the frequency of switching common to all the edges,
and $\{ \theta_{ij} \}$ for $i$--$j$ pairs with $a_{ij} = 1$ is the set of edges' intrinsic phases, which is
randomly assigned from the uniform distribution of the interval
$[0, \pi )$. By choosing such an interval, the edge densities
change periodically with the exact triangle wave subject to the frequency $\Omega$,
from fully connected edges to isolated oscillators without interactions.
Note that the factor $2$ in Eq.~(\ref{Kuramoto_equation}) is to directly compare the coupling
strength $K$ of our model
with $\overline{T_{ij}} = 1/2$
to the $K$ value in the original Kuramoto model,
where $\overline{x}$ refers to the temporal average of $x(t)$ over
a period.

In addition to the original control parameter of coupling strength
$K$ in the Kuramoto model, the edge frequency $\Omega$ driving the system
is a crucial parameter as well. The response of oscillators to the toggling
interactions is measured by the temporal behavior of conventional phase order
parameter $\Delta(t)$ where
\begin{equation}
\Delta(t) \exp[i\psi(t)] = \frac{1}{N} \sum_{i=1}^N \exp[i\phi_{i}(t)] .
\label{phase_order_parameter}
\end{equation}
The order parameter's dynamics depends on the periodic change
of interactions with the density
$\rho(t) = \sum_{i<j} a_{ij} T_{ij}(t) / \sum_{i<j} a_{ij}$.
We provide the java applet simulating this model system in case of globally connected oscillators,
for readers to observe the behavior of a system of relatively small sizes~\cite{JavaApplet}.
From now on, we present our numerical simulation results and their
implications.

\section{results}
\label{simulation}

\subsection{Globally coupled oscillators}
\label{sec:all_to_all_case}

\begin{figure}
\includegraphics[width=0.9\columnwidth]{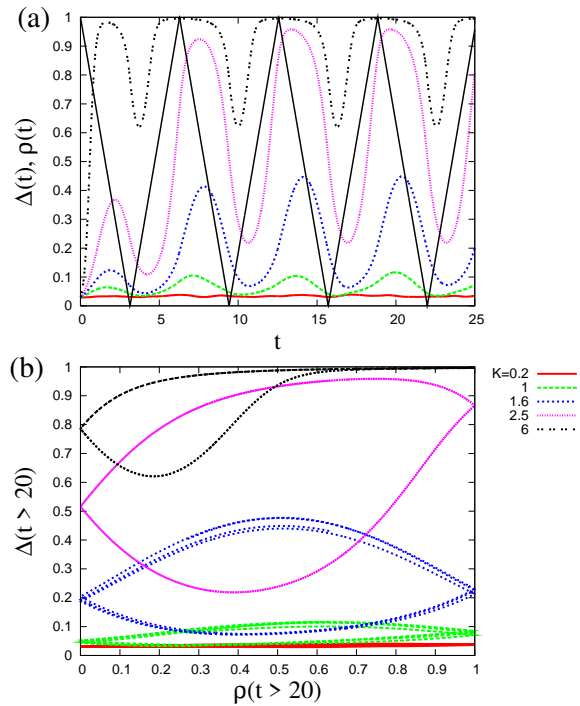}
\caption{A typical behavior of the order parameter
$\Delta(t)$ and the interaction density $\rho(t)$. The system size
$N = 800$, the edge frequency $\Omega = 1$, and the time series of
$\Delta(t)$ depending on $K$ are shown as smooth curves in (a) and
$\rho(t)$ (independent of $K$) as black triangle waves in (a). The
$\Delta$--$\rho$ diagram for $t > 20$ after the transient behavior
is shown in (b). All the curves are results averaged over $50$
samples.
}
\label{typical_example}
\end{figure}

\begin{figure}
\includegraphics[width=0.8\columnwidth]{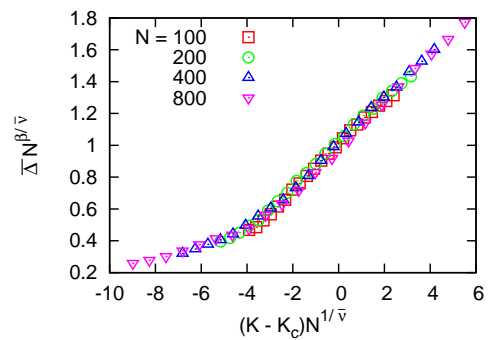}
\caption{FSS scaling collapse of the temporally averaged order parameter
$\overline{\Delta}$ for
various system sizes with the critical exponents of the original MF Kuramoto model
[$K_c = 1.62(1)$, $\beta = 1/2$, and $\bar{\nu} = 5/2$]. The edge
frequency is $\Omega = 1$.
}
\label{FSS}
\end{figure}

First, we consider the globally coupled case, i.e., $a_{ij} = 1 - \delta_{ij}$.
A typical temporal behavior of oscillators is illustrated in Fig.~\ref{typical_example}, in case of $\Omega = 1$.
Note that the oscillating behavior of the order parameter $\Delta(t)$
driven by switching edges (with the same frequency)
is shown even in case of quite low values of $K$,
for which the original MF Kuramoto model corresponds to the disordered phase.
Another interesting aspect is the relative phase shift of $\Delta(t)$
with respect to $\rho(t)$. In other words, the maximum or minimum
of $\Delta(t)$ occurs after some time when the maximum of minimum
of $\rho(t)$ is reached, as shown in Fig.~\ref{typical_example}(a).
Moreover, as $K$ is increased, the delay for maximum values and
the delay for minimum values become more asymmetric, as clearly
shown in Fig.~\ref{typical_example}(b).
In spite of such temporal oscillations of $\Delta(t)$, the temporally
averaged value $\overline{\Delta}$ (averaged after the transient time $\simeq 20$)
shows exactly the same universality class of synchronization transition as the MF Kuramoto model with
$\beta = 1/2$, and $\bar{\nu} = 5/2$~\cite{Acebron2005,HHong2007},
as described in the FSS scaling collapse with $\overline{\Delta} = N^{- \beta / \bar{\nu}} f \left( \left(K - K_c (\Omega)\right) N^{1/{\bar{\nu}}} ; \Omega \right)$
(the scaling function $f$ itself, as well as the critical coupling strength $K_c$, depends on $\Omega$)
shown in Fig.~\ref{FSS}. We suggest the slightly larger value of $K_c (\Omega)$ than that of the MF model $K_c^{\textrm{MF}} = 2\sqrt{2/\pi} \simeq 1.596$~\cite{Acebron2005} stem from additional quenched randomness caused by $\{ \theta_{ij} \}$ for edges,
similarly to the lower $T_c$ (the disordered phase extended) for the kinetic Ising model under the external oscillating field~\cite{BJKim2001}.
Since we have confirmed the MF transition for $\overline{\Delta}$, from now on we focus the temporal
behavior of $\Delta(t)$, especially with respect to the oscillating interaction strength $\rho(t)$.

\begin{figure}
\includegraphics[width=0.90\columnwidth]{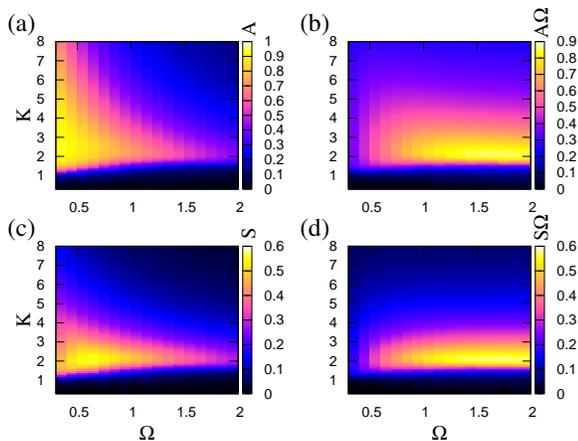}
\caption{The difference $A$ between maximum and minimum
values of $\Delta(t)$ oscillation and the area $S$ enclosed by the loop
in $\Delta$--$\rho$ the diagram are shown in the $K$--$\Omega$ plane.
The system size
$N = 800$, and the color-coded values are $A$ (a), $A \Omega$ (b),
$S$ (c), and $S \Omega$ (d).
The time series $\Delta(t)$ for $t > 20$ after the transient behavior
is used, and all the results are averaged over $50$
samples.
}
\label{amplitude_area_density}
\end{figure}


\begin{figure}
\includegraphics[width=\columnwidth]{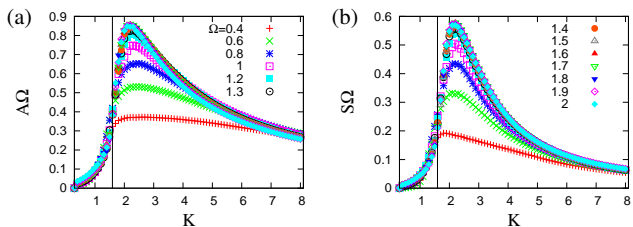}
\caption{$A\Omega$ (a) and $S\Omega$ (b) as functions
of $K$, for various $\Omega$ values. The black vertical lines correspond to the MF
critical coupling strength $K_c^{\textrm{MF}} = 2\sqrt{2/\pi}$.
}
\label{amplitude_area_2D}
\end{figure}

To characterize the oscillating behavior more quantitatively, we
systematically measure the difference between maximum and minimum values of $\Delta(t)$
oscillation denoted as $A$ (roughly twice the amplitude of $\Delta(t)$) and the enclosed area
in the $\Delta$--$\rho$ diagram denoted as $S$, for a wide range of
$\Omega$ and $K$ values. 
Note that the data shown here are from
the system size $N = 800$, but we have checked that there exists no significant
finite-size effects based on the simulations results with smaller and
larger systems. The time series $\Delta(t)$ for $t > 20$ after the transient behavior
is used, and all the results are from averaged over $50$
samples for all the simulations from now on.
As shown in Fig.~\ref{amplitude_area_density}, the scaling behavior of $A$ and
$S$ is quite similar, and notably both $A$ and $S$
seem to be inversely proportional to $\Omega$ for $\Omega \gtrsim 1.2$.
This scaling behavior makes the $A\Omega$ and $S\Omega$ in Figs.~\ref{amplitude_area_density}(b) and
(d) depend only on $K$ for $\Omega \gtrsim 1.2$, and is also clearly
observable from the fact that the curves for $\Omega \gtrsim 1.2$ are collapsed
in Fig.~\ref{amplitude_area_2D}. Since $\Omega$ represents the angular velocity
of the fluctuation of edges (hence that of $\Delta(t)$), $A \Omega$ corresponds to
the ``linear velocity'' of $\Delta(t)$, which is shown to be conserved for a given
value of $K$. More importantly, the synchronization transition is clearly
shown in the steep change of $A$ and $S$ near
the critical coupling strength $K_c^{\textrm{MF}}$ for the original MF Kuramoto
model, as shown in Figs.~\ref{amplitude_area_density}
and \ref{amplitude_area_2D}. Therefore, we conclude that the signature of the
MF synchronization transition of Kuramoto model is resurfaced
in terms of $A$ and $S$.

Aside from the amplitude of $\Delta$ and enclosed area in the $\Delta$--$\rho$
diagram, a notable property of the $\Delta$--$\rho$ diagram in Fig.~\ref{typical_example}(b)
is successive change in the shapes of enclosed regions, as $K$ is increased.
First, the desynchronization process of $\Delta(t)$ after the onset of decreasing phase
of $\rho(t)$ is slower than the synchronization process of $\Delta(t)$ after the onset of
increasing phase of $\rho(t)$ in general.
This asymmetry is natural,
since the response of the order parameter
to such linear functions is integrated form, i.e., quadratic functions depending
on the initial values.



\begin{figure}
\includegraphics[width=0.95\columnwidth]{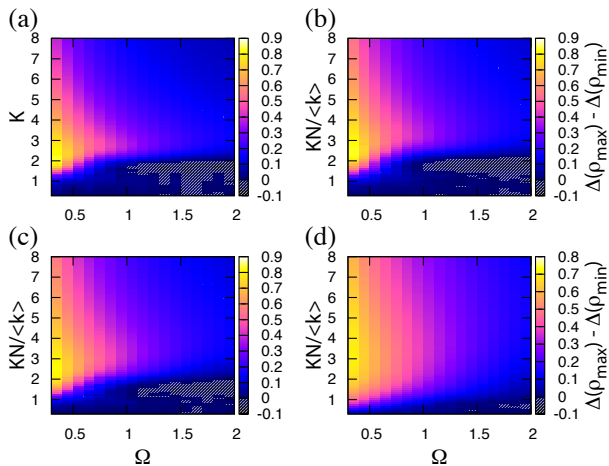}
\caption{Height difference between $\Delta(\rho_\textrm{max})$
and $\Delta(\rho_\textrm{min})$, for the globally coupled
case (a) and SFNs with $\gamma = 6.5$ (b), $3.75$ (c), and $2.2$ (d).
Note that the rescaled coupling strength $KN/\langle k \rangle$, where $\langle k \rangle$ is set as $20$,
is used for (b)--(d), to compare with the globally coupled case (a) with the same
Eq.~(\ref{Kuramoto_equation}). The same system size $N = 800$ is used for all the cases.
}
\label{height_diff}
\end{figure}

We show that the synchronization transition is also observed in the phase shift of $\Delta(t)$
with respect to $\rho(t)$ as well.
To systematically analyze such a phase shift caused by the temporal delay of $\Delta(t)$,
we check the relative height difference of $\Delta$ at the right and left ends in Fig.~\ref{typical_example}(b),
corresponding to $\Delta(\rho_\textrm{max})$ and $\Delta(\rho_\textrm{min})$ respectively,
also affects the shape of enclosed regions.
In Fig.~\ref{height_diff}(a),
we plot such height differences in the $K$--$\Omega$ space, and the height difference
occurs mainly in small $\Omega$ and large $K$ regimes. Interestingly, near $K \simeq 1.6$
in relatively large $\Omega$ regimes, the difference becomes negative, which corresponds to
$\Delta(\rho_\textrm{max}) < \Delta(\rho_\textrm{min})$. We call this phenomenon {\em phase-shift
inversion}. $K \simeq 1.6$ is close to the MF critical
coupling constant $K_c^{\textrm{MF}}$ again,
and we suggest that it is related to critical slowing down near the critical
point in a phase transition, as shown in the diverging relaxation time $\tau$ at $K_c^{\textrm{MF}}$~\cite{Strogatz1991,SWSon2008},
given by
$1 = \sqrt{\pi/8} K \exp[1/(2\tau^2)] \textrm{erfc}[1/(\sqrt{2}\tau)]$.
This resonance-like phase-shift inversion phenomenon occurs similarly to the
frequency matching condition around the phase transition in the MF kinetic
Ising model~\cite{BJKim2001}. In other words, the MF synchronization transition
in the Kuramoto model is also reflected in the temporal response to the periodically
switching interactions.

\subsection{Scale-Free Networks}
\label{sec:SFN}

To validate our conclusion on the phase transition in terms of phase-shift inversion,
we use substrate network topologies other than the globally coupled one. The
static scale-free network (SFN) model~\cite{KIGoh2001} with the tunable degree distribution
$p(k) \sim k^{-\gamma}$ provides various underlying interaction structures such as
the pure MF case ($\gamma > 5$), the MF with critical exponents
depend on $\gamma$ ($3 < \gamma < 5$), and eventually the one where critical coupling strength
vanishes as $K_c \to 0$ ($\gamma < 3$)~\cite{HHong2007}. We use the SFN model for $\{ a_{ij} \}$ in Eq.~(\ref{Kuramoto_equation})
and observe the phase-shift inversion
diagram for the three representative cases: $\gamma = 6.5$, $3.75$, and $2.2$, as shown in
Figs.~\ref{height_diff}(b)--(d).
One can observe the phase-shift inversion in SFN cases where
the synchronization transition exists ($\gamma > 3$) as well, at lower coupling strength
explained by $K_c \propto 1/\langle k^2 \rangle$ when $\langle k \rangle$ is fixed~\cite{HHong2007},
where $\langle x \rangle$ represents the averaged $x$ over the entire nodes in a network.
The $\gamma < 3$ case
reveals its vanishing critical point in the fact that the seeming phase-shift inversion
occurs in much lower values (and over the very narrow region) of coupling strength [Fig.~\ref{height_diff}(d)],
due to the finite-size effect. Therefore, the results from SFNs also clearly support
the fact that phase-shift inversion reflects the phase transition.

\section{summary and discussions}
\label{summary}
We have studied the Kuramoto model under periodically switching interactions,
checked that it shows the same mean-field synchronization transition as the
original Kuramoto model with the finite-size scaling analysis,
and found that the phase transition is observed in terms of dynamical properties
such as the amplitude of oscillation of the order parameter and
its relative phase shift with respect to the overall strength of
interactions. In particular, the latter causes the phase-shift inversion
phenomenon that the significantly large phase shift near the phase transition let
the order parameter at the minimum interaction density to be larger than
that at the maximum interaction density. The conclusion does not only hold for the globally connected
substrate structure, but also holds for other types of substrate
structures such as scale-free networks. Such observations strongly suggest that our model
is related to the mean-field kinetic Ising model with the frequency matching condition around
the phase transition~\cite{BJKim2001} and would be worth investigating further.
Since periodicity plays important roles in many parts of the nature, especially for systems
under natural circadian rhythms
or external stimuli, our results suggest that the temporal response to such periodicity
can be an important cue to characterize the system.

\begin{acknowledgments}
We greatly appreciate B.~J. Kim, H.-H. Jo, M. Gastner, T. Gross, and S. Dorogovtsev
for their valuable comments. This work was supported by the Swedish Research Council (S.H.L. and P.H.),
the Wenner-Gren Foundation (S.L.),
and the WCU program through NRF Korea funded by MEST R31-2008-10029 (P.H.).
Computation was
partially carried out using the cluster in CSSPL, KAIST.
\end{acknowledgments}


\end{document}